\documentclass[10pt]{article} 
\usepackage{latexsym,amsmath,amssymb}
\usepackage{lscape,enumitem}
\usepackage{epstopdf,caption,subcaption,graphicx,xcolor,algorithm,algorithmicx,algpseudocode}
\usepackage{tikz,pgfgantt,pgfplots,needspace}
\usepackage{hyperref}
\usepackage[capitalise,noabbrev]{cleveref}
\usetikzlibrary{arrows.meta,positioning,fit,backgrounds,calc}
\usetikzlibrary{shapes,arrows,fit,calc,positioning}
\usetikzlibrary{decorations.pathreplacing}
\normalsize
\newtheorem{theorem}{Theorem}
\newtheorem{lemma}{Lemma}
\newtheorem{corollary}{Corollary}

\newenvironment{proof}{{\sc Proof. }}{\hfill$\Box$\vspace{0.1in}}
\newcommand{\OPT}{{\rm OPT}}
\newcommand{\cost}{{\rm cost}}
\newcommand{\dist}{{\rm dist}}
\newcommand{\poly}{{\rm poly}}
\newcommand{\Split}{{\rm Split}}

\newcommand{\E}{{\mathbb E}}

\def\mcA{\mathcal{A}}

\def\mcC{\mathcal{C}}
\def\mcF{\mathcal{F}}
\def\mcH{\mathcal{H}}
\def\mcJ{\mathcal{J}}
\def\mcO{\mathcal{O}}
\def\mcP{\mathcal{P}}
\def\mcQ{\mathcal{Q}}

\def\mcR{\mathcal{R}}

\title{Improved approximation algorithms for capacitated cluster editing}
\author{
Cong~Ma\thanks{School of Mathematics, Hangzhou Dianzi University. Hangzhou, China.
	Emails: \texttt{\{244070121, chenyong, 245070122\} @hdu.edu.cn}}
\and
Guangting~Chen\thanks{Zhejiang University of Water Resources and Electric Power. Hangzhou, China.
	Email: \texttt{gtchen@hdu.edu.cn}; ORCID: \texttt{0000-0003-3013-3325}}
\and
Yong~Chen$^*$\thanks{Corresponding authors.}
	\thanks{ORCID: \texttt{0000-0001-8982-7757}}
\and
Guohui~Lin\thanks{Department of Computing Science, University of Alberta.  Edmonton, Canada.
	Email: \texttt{guohui@ualberta.ca}; ORCID: \texttt{0000-0003-4283-3396}} $^{\,\ddagger}$
\and
Jiaxuan~Ma$^*$
\and
Lusheng~Wang\thanks{Department of Computer Science, City University of Hong Kong. Hong Kong, China.
	Email: \texttt{cswangl@cityu.edu.hk}; ORCID: \texttt{0000-0002-4344-8791}}
}
\date{}
\begin{document}
\maketitle

\begin{abstract}
We study the minimum-disagreement cluster editing problem that,
from a complete edge-signed graph, where a positive edge represents two vertices agreeing to each other while a negative represents a disagreement,
constructs disjoint cliques of upper bounded size such that the total number of negative edges inside cliques and positive edges outside cliques is minimized.
When the sizes of the candidate cliques are unbounded,
the problem is NP-hard and APX-hard, and many approximation algorithms with improving ratios have been proposed in the literature,
with the most recent ones appearing in {\em JACM 2008}, {\em STOC 2015}, {\em FOCS 2022}, {\em FOCS 2023}, {\em STOC 2024} and {\em arXiv 2026}
and the state-of-the-art approximation ratio of $\alpha = 1.3865 + \epsilon$ for any $\epsilon > 0$.
The capacitated variant, in which the candidate cliques have sizes no greater than a given $k$,
is less studied ({\em SIAM J. Optim 2015}, {\em IJFCS 2024}) and the best approximation ratio is $5.37$.
In this paper, we study the latter problem from the perspective of approximation algorithms.
We first present a downsizing procedure that achieves an approximation ratio of $\alpha_k = 2 \alpha + 1$,
based on any $\alpha$-approximation algorithm for the uncapacitated problem.
Secondly, we propose a hybrid of cluster-LP sampling and pivoting, followed by the downsizing procedure,
and show that for every fractional capacitated cluster-LP solution of cost $L$, the hybrid using ordinary pivoting has an expected cost at most $\frac 95 L + \frac 1{10} \OPT_k$,
and the hybrid using conditional pivoting has an expected cost at most $\frac {306}{203} L + \frac {50}{203} \OPT_k$.
A cardinality-constrained adaptation of the approximate dual-separation framework of Garc\'ia-Soriano and Schohn
supplies the required fractional solution and yields a randomized $(1.9 + \epsilon)$-approximation algorithm
and a randomized $(1.7537 + \epsilon)$-approximation algorithm, respectively, for any $\epsilon > 0$.
All these three algorithms have running time polynomial in the size of the graph, the number of positive values in the fractional solution, and the encoding length,
and the exponents are independent of $k$.

\end{abstract}

\section*{Acknowledgments}
The research is supported by the NNSF of China (Grants 12471301, 61972329),
the China Scholarship Council (Grant No. 202508330173),
the GRF grants for Hong Kong Special Administrative Region, China (Grant Nos. CityU 11218423, CityU 11219825), and
the NSERC Canada.

\newpage
\section{Introduction}
Correlation clustering, introduced by Bansal et al.~\cite{BBC04}, constructs a partition from pairwise similarity and dissimilarity information.
In the complete unweighted form, every pair of vertices is labelled positive or negative.
The objective is to form a partition of the vertices, in which a part is called a cluster, to minimize the number of positive edges between clusters and negative edges within clusters.
Equivalently, one seeks to transform the graph of positive edges only into a disjoint union of cliques using as few edge insertions and deletions as possible.
Such a formulation is also known as {\em Cluster Editing}, which is an NP-hard~\cite{SST04} and APX-hard~\cite{CGW05} problem.

We study {\em Capacitated} Cluster Editing, in which each candidate cluster contains no more than $k$ vertices, where $k$ is part of the input and is referred to as the cluster capacity.
The problem remains NP-hard for every fixed $k \ge 3$~\cite{IN11}.
On the approximation algorithms, Puleo and Milenkovic~\cite{PM15} developed an algorithm based on a linear program and combinatorial methods,
including two proven $7$- and $6$-approximation algorithms.
Il'ev et al.~\cite{IIN16} established the APX-completeness for each fixed capacity, with an approximation ratio depending on that capacity.
Later, Ji et al.~\cite{JCT24} improved the approximation ratio to $5.37$, independent on the capacity, based on a linear program formulation and rounding to construct size-bounded clusters.
Kononov and Il'ev~\cite{KI25} further obtained better ratios of $\frac 32$ for $k = 3$ and $\frac {13}7$ for $k = 4$.
In this paper, we aim to design approximation algorithms with improved ratios when $k$ is part of the input.

The problem introduced by Bansal et al.~\cite{BBC04} is uncapacitated, i.e., a candidate cluster can have arbitrary size.
Bansal et al.~\cite{BBC04} had presented a constant ratio approximation algorithm, with the ratio being at most $9 \cdot 44^2 + 5 = 17,429$.
Recent progress for the uncapacitated problem has been driven by the {\em Cluster} linear program,
which represents a fractional partition by assigning weights to all candidate clusters~\cite{CCL24,GS26}.
The same representation can employed for capacitated variant,
and in fact it makes an upper bound on cluster size easy to encode, but subsequently rounding poses a further difficulty.
Basically, from a fractional solution, sampling the clusters preserves the size bound, whereas pivoting can create oversized clusters;
for the latter oversized clusters, the feasibility can be restored by downsizing which however may increase the disagreement cost.
Obtaining strong guarantees therefore requires incorporating downsizing into the rounding analysis.

\subsection{Our results}
We first propose a simple downsizing procedure that converts any uncapacitated partition $\mcP$ into a feasible capacitated partition denoted as $\Split(\mcP)$
having a disagreement cost at most twice of $\mcP$ plus an optimum.
This gives rise to a $(3.7730 + \epsilon)$-approximation algorithm based on the state-of-the-art $(1.3865 + \epsilon)$-approximation for the uncapacitated Cluster Editing problem.

Next, realizing that we can extend the uncapacitated Cluster-LP to capacitated Cluster-LP to achieve a fractional solution of cost arbitrarily close to the optimum,
we propose to combine the cluster sampling approach of Cao et al.~\cite{CCL24} to obtain a feasible solution $\mcR$,
and the {\em ordinary pivot} method introduced by Ailon et al.~\cite{ACN08} to obtain another uncapacitated solution $\mcP$,
so that the better solution between $\mcR$ and $\Split(\mcP)$ is a $(1.9 + \epsilon)$-approximation.

Lastly, we propose to employ the conditional-affine pivoting framework of Garc\'ia-Soriano and Schohn~\cite{GS26} with different vertex admission probabilities for
positive neighbors inside the seed, positive neighbors outside the seed, and negative neighbors.
Coupling with a weighted disagreement cost analysis, which calculates more accurately the edge errors and thus estimates slightly better than the downsizing argument,
we are able to show that the new pivoting method performs better than the ordinary pivoting,
and the better solution between $\mcR$ and $\Split(\mcP)$ is a $(1.7537 + \epsilon)$-approximation.

\begin{theorem}
\label{thm01}
For every fixed rational $\epsilon > 0$, the Capacitated Cluster Editing problem admits a randomized expected $(\frac {356}{203} + \epsilon)$-approximation algorithm
that runs in an expected polynomial time in the input size $n$, bounded by $n^{\poly(1/\epsilon)}$ in which the exponent does not depend on the capacity $k$,
times a polynomial in the precision encoding length.
\end{theorem}

Note that $\frac {356}{203} = 1.753694581280\dots$.
Our main contribution is to combine the rounding methods proposed in the literature with the downsizing procedure and analyze their costs jointly.
The improvement from the first pivoting method to the second comes from a better bound on the edge errors together with a weighted disagreement cost analysis.

\subsection{Related work}
Approximation algorithms for uncapacitated correlation clustering have evolved through increasingly expressive linear program (LP) relaxations and rounding procedures.
Improving upon Bansal et al.~\cite{BBC04}'s large constant ratio, Charikar et al.~\cite{CGW05} gave a $4$-approximation using LP rounding and region growing.
Ailon et al.~\cite{ACN08} introduced randomized {\em pivoting} method, which repeatedly groups a random unclustered vertex with its positive neighbors, obtaining a $3$-approximation.
Furthermore, their LP-dependent rounding achieves a ratio of $2.5$.
Van Zuylen and Williamson~\cite{VW09} also presented deterministic pivoting algorithms with the same two guarantees.
Afterwards, Chawla et al.~\cite{CMS15} refined the LP-based rounding probabilities in \cite{ACN08} to obtain an improved ratio slightly below $2.06$.
It is proven that the underlying metric LP using pairwise separation variables constrained by triangle inequalities has an integrality gap lower bound of $2$~\cite{CGW05,CMS15}.

In the past few years, uncapacitated correlation clustering has experienced another wave of advancement,
mostly stronger relaxations retaining correlations among several clustering decisions.
Cohen-Addad et al.~\cite{CLN22} combined the Sherali-Adams hierarchy with correlated rounding and a global charging argument to obtain a $(1.994 + \epsilon)$-approximation algorithm.
Cohen-Addad et al.~\cite{CLL23} introduced pre-clustering, which identifies groups of vertices to be kept together and restricts those pairs to share a cluster;
the proposed structure controls correlated-rounding errors and supports a combination of set-based and pivot-based rounding with a guarantee of $1.73 + \epsilon$.

The cluster LP of Cao et al.~\cite{CCL24} brings the above ideas into a single relaxation over all candidate clusters,
by setting up whole-cluster variables to retain joint membership information beyond pairwise distances.
Their fractional partition construction uses pre-clustering and sampling to obtain an explicit feasible solution to the cluster LP of value
no more than a fraction $\epsilon$ of the integral optimum.
Subsequently, correlated rounding errors are handled so that the final rounding can work directly with the resulting cluster weights.
Combining cluster sampling with LP-dependent pivoting gives a $(1.485 + \epsilon)$-approximation algorithm in their extended full paper~\cite{CCL25}.

Soon after, Garc\'ia-Soriano and Schohn~\cite{GS26} improved both the computation and the rounding of the cluster LP.
Their approximate dual-separation algorithm uses localization and weak regularity to search for violated dual constraints,
yielding approximate primal and dual solutions relative to the fractional cluster-LP optimum.
Their rounding couples cluster sampling with a conditional affine pivot rule, which uses membership in a sampled cluster to coordinate inclusion around the selected pivot.
A variance-based analysis and derandomization yield a deterministic $(1.3865 + \epsilon)$-approximation.
This second wave of developments supply the cluster LP optimization framework and rounding procedures that we adapt to for our capacitated Cluster Editing.

Besides the above mentioned constant-ratio approximation algorithms for the capacitated Cluster Editing~\cite{PM15,IIN16,JCT24,KI25},
Ji et al.~\cite{JXL19} consider a variant in which vertex-specific bounds are imposed on the size of the cluster containing each vertex, i.e., non-uniform capacity constraints,
and they presented a bicriteria algorithm achieving a cost at most $4$ times the optimum while allowing a factor-$2$ violation of these bounds.
In this paper, we study the problem with prescribed uniform capacity.

Cluster Editing has also been studied from the perspectives of exact and parameterized algorithms.
For example, the problem can be parameterized by the edit-cost budget, which has received substantial studies~\cite{BBB09,Guo09,Dam10,BD11,CC12,CM12},
and by jointly the edit-cost budget and the number of output clusters~\cite{FKP14}.
Practical approaches for uncapacitated Cluster Editing include those that combine data reduction with exact branching and integer programming~\cite{BBK11},
and that integrate exact optimization with greedy randomized adaptive search and iterated-local-search heuristics~\cite{BOP16}.

\subsection{Organization of the paper}
In the next section, we introduce a set of terminologies and notations to be used in the rest of the paper, as well as the capacitated Cluster LP formulation.
The downsizing procedure is presented in Section~3, followed by what it implies ---
our first $(3.7730 + \epsilon)$-approximation algorithm building on the state-of-the-art algorithm for uncapacitated problem.
In Section~4, we employ the near optimal fractional solution to the Cluster LP, which can be computed in polynomial time,
and combine the rounding through sampling by Cao et al.~\cite{CCL24} and the ordinary pivoting method by Ailon et al.~\cite{ACN08}, together with the downsizing procedure,
to design an improved approximation algorithm with ratio of $19/10 + \epsilon$.
We then replace the ordinary pivoting method by the conditional affine pivoting rule of Garc\'ia-Soriano and Schohn~\cite{GS26} to
construct another uncapacitated partition dependent on the fractional solution to the Cluster LP, again together with the downsizing procedure,
leading to a further improved $(356/203 + \epsilon)$-approximation algorithm, where $356/203 = 1.753694581280\dots$.
These results are presented in Section~5.
We conclude the paper in Section~6.

\section{Preliminaries}\label{sec:prelim}
In this paper, we study the {\em Cluster Editing} problem.
An instance of the problem consists of a complete edge-signed graph $G =(V, E^+ \cup E^-)$,
where an edge $uv \in E^+$ is called a {\em positive} edge or otherwise $uv \in E^-$ a {\em negative} edge,
and a positive integer $k$ when the problem is {\em capacitated} or otherwise $k = \infty$ when the problem is {\em uncapacitated}.
A partition $\mcP$ of the vertex set $V$ is a collection of disjoint nonempty subsets (called {\em parts} or {\em clusters}, used interchangeably) whose union is $V$.
For the uncapacitated variant there is no constraint on the size of a part in the partition;
for the capacitated variant, every part must have size at most $k$ for the partition to be {\em feasible} (called $k$-feasible).
This parameter $k$ is referred to as the (uniform) {\em capacity} of the clusters.
Note that there is no requirement on the number of parts in a partition.

For a feasible partition $\mcP$, the edge set of $\mcP$ is denoted as $	I(\mcP)$,
which contains all the (positive and negative) edges in the complete subgraphs (called {\em cliques}) induced by the parts of $\mcP$,
i.e., those edges $uv$ such that $u, v$ lie in the same part of $\mcP$.
The {\em disagreement cost} of $\mcP$ is
\begin{equation}
\label{eq01}
\cost(\mcP) = |E^+ \setminus I(\mcP)| + |E^- \cap I(\mcP)| = |E^+ \setminus I(\mcP)| + |I(\mcP) \setminus E^+| = \dist(E^+, I(\mcP)),
\end{equation}
which is the {\em edit distance} from the positive edge set $E^+$ to $I(\mcP)$, by deleting certain edges of $E^+$ and inserting certain edges of $E^-$.
The goal of the Cluster Editing problem is to find a feasible partition that minimizes the cost.
We use $\OPT_k$ and $\OPT_{\infty}$ to denote the optimal (i.e., minimum) cost for the $k$-capacitated and the uncapacitated variants, respectively.
One sees that $\OPT_\infty \le \OPT_k$.

In general, for two edge subsets $E_1$ and $E_2$, their {\em symmetric difference} is $E_1 \triangle E_2 = (E_1 \setminus E_2) \cup (E_2 \setminus E_1)$,
and their edit distance is $\dist(E_1, E_2) = |E_1 \triangle E_2|$.
One sees that $\dist()$ is a metric.

Below we give another view for the disagreement cost of a partition $\mcP$.
For two disjoint vertex subsets $P_1$ and $P_2$, let $e^+(P_1, P_2)$ count the number of positive edges with one endpoint in each set;
let $e^+(P_1)$ and $e^-(P_1)$ count the number of positive and negative edges in the clique induced on $P_1$, respectively.
The {\em cost} of using a vertex subset $S$ as a part in a partition is
\begin{equation}
\label{eq02}
c(S) = e^-(S) + \tfrac 12 e^+(S, V \setminus S),
\end{equation}
in which $e^-(S)$ represents the number of insertions due to $S$ and $\frac 12 e^+(S, V \setminus S)$ the number of deletions.
Note that the other half of the deletions is attributed to the parts containing the other endpoints of the edges of $e^+(S, V \setminus S)$.
This way, we have $\cost(\mcP) = \sum_{P \in \mcP} c(P)$.

For a vertex $v \in V$, if $uv \in E^+$ then $u$ is a {\em positive neighbor} of $v$;
the {\em positive neighborhood} of $v$ is $N^+(v)$ containing all its positive neighbors,
the {\em closed positive neighborhood} of $v$ is $N^+[v] = \{v\} \cup N^+(v)$, and
the {\em positive degree} of $v$ in a vertex subset $S \subseteq V$ is $d_S^+(v) = |N^+(v) \cap S|$.
We write $d^+(v) = d^+_V(v)$.

\subsection{The capacitated cluster linear program}
Let $\mcF_k = \{S \subseteq V : 1 \le |S| \le k\}$, i.e., all the non-empty vertex subsets of size up to $k$.
For any pair of distinct vertices $u, v$, we denote $\mcC(u, v) = \{S \in \mcF_k \mid \{u, v\} \subseteq S\}$, i.e., the collection of all subsets in $\mcF_k$ containing both vertices.
We also denote $\mcC(v) = \{S \in \mcF_k \mid v \in S\}$ for any vertex $v$.
Using the definition of cluster cost in Eq.~\eqref{eq02},
we set up a binary variable $z_S$ for including $S$ as a part in the partition, followed by the {\em capacitated cluster} relaxed linear program (abbreviated as Cluster-LP):
\begin{align}
L_k^* = \min\quad 	&\sum_{S \in \mcF_k} c(S) z_S\label{eq03}\\
\mbox{s.t. }\quad 	&\sum_{S \in \mcC(v)} z_S = 1, \ \forall v \in V;\nonumber\\
		&z_S \ge 0, \ \forall S \in \mcF_k.\nonumber
\end{align}
Here, $\sum_{S \in \mcC(v)} z_S = 1$ is called the {\em coverage equality} for the vertex $v$.
As $L^*_k$ denotes the optimal disagreement for the Cluster-LP, we have $L_k^* \le \OPT_k$.

Given a feasible solution $z$ to the Cluster-LP, we define for each edge $uv$,
\begin{equation}
\label{eq04}
y_{uv} = \sum_{S \in \mcC(u, v)} z_S, \mbox{ } x_{uv} = 1 - y_{uv}.
\end{equation}
The vertex coverage equalities imply $0 \le y_{uv} \le 1$, and hence $0 \le x_{uv} \le 1$ too.
For a positive edge $uv$, the total disagreement across all the clusters containing exactly one endpoint is
$\sum_{S \in \mcC(u)} z_S + \sum_{S \in \mcC(v)} z_S - 2 y_{uv} = 2 - 2 y_{uv} = 2 x_{uv}$;
therefore, by Eq.~\eqref{eq02} this edge {\em contributes} $x_{uv}$ to the objective.
Similarly, for a negative edge $uv$, the total disagreement across all the clusters containing both endpoints is $y_{uv} = 1 - x_{uv}$.
It follows that the objective function value of the solution $z$ can be written in three ways as
\[
L(z) = \sum_{S \in \mcF_k} c(S) z_S = \sum_{uv \in E^+} x_{uv} + \sum_{uv \in E^-} (1 - x_{uv}) = \sum_{uv \in E^+} (1 - y_{uv}) + \sum_{uv \in E^-} y_{uv}.
\]

For any three distinct vertices $u, v$ and $w$, since $\mcC(u, w) \cup \mcC(v, w) \setminus \mcC(u, v) \subseteq \mcC(w)$ and $z_S \ge 0$ for all $S \in \mcF_k$,
we have $y_{uw} + y_{vw} - y_{uv} \le 1$, and consequently the triangle inequality holds for $x$
\begin{equation}
\label{eq05}
x_{uv} \le x_{uw} + x_{vw}, \ \forall \mbox{distinct } u, v, w.
\end{equation}

\section{A downsizing theorem}\label{sec:transfer}
Given a partition $\mcP$, we want to downsize every part of size greater than $k$ to convert it into a $k$-feasible solution to the capacitated problem.
We do this by splitting each part of size $q k + r$, where $q \ge 1$ and $0 \le r < k$, into $q$ subsets of size exactly $k$ and a subset of size $r$ if $r > 0$.
Note that there is no requirement on co-memberships for any vertices during the splitting.
The achieved partition through such a procedure is denoted as $\Split_k(\mcP)$.

\begin{lemma}
\label{lemma01}
Given a partition $\mcP$, $\dist(I(\mcP), I(\Split_k(\mcP))) \le \dist(I(\mcP), I(\mcA))$ for any $k$-feasible partition $\mcA$.
\end{lemma}
\begin{proof}
We use the $k$-feasible partition $\mcA$ to intersect the parts of $\mcP$ to obtain a common refinement denoted as $\mcA \wedge \mcP$.
In more details, let $A$ be a part of $\mcA$ and $P$ be a part of $\mcP$;
then $A \cap P$ becomes a part in the refinement if it is non-empty.
$\mcA \wedge \mcP$ is $k$-feasible and its edge set $I(\mcA \wedge \mcP) = I(\mcA) \cap I(\mcP)$.
It follows that
\[
\dist(I(\mcP), I(\mcA \wedge \mcP)) = |I(\mcP) \setminus I(\mcA)| \le \dist(I(\mcP), I(\mcA)).
\]
It thus suffices to prove the lemma for only the refinements of $\mcP$, and furthermore independently for each part of $\mcP$.

Let $P$ be a part of $\mcP$.
If a $k$-feasible refinement of $P$ contains two subsets of sizes $0 < s_1 \le s_2 < k$,
then transferring them into a subset of size $s_2 + 1$ and, when $s_1 > 1$, another of size $s_1 - 1$,
gives rise to a better refinement reducing the distance by $s_2 - s_1 + 1 > 0$.
Repeating such a process leads to the $k$-feasible partition $\Split_k(\mcP)$.
This proves the lemma.
\end{proof}

\begin{theorem}[Downsizing]
\label{thm02}
For every partition $\mcP$ of the complete edge-signed graph $G = (V, E^+ \cup E^-)$, $\cost(\Split_k(\mcP)) \le 2 \cdot \cost(\mcP) + \OPT_k$.
Hence a polynomial-time $\alpha$-approximation algorithm for the uncapacitated cluster editing problem yields
a polynomial-time $(2 \alpha + 1)$-approximation algorithm for the capacitated problem.
The same implication holds for performance guarantees in expectation.
\end{theorem}
\begin{proof}
Let $\mcO$ be an optimal $k$-feasible partition for the cluster editing problem with capacity $k$ and $\mcQ = \Split_k(\mcP)$.
By Eq.~\eqref{eq01}, the triangle inequality for the edge set edit distance $\dist()$ and Lemma~\ref{lemma01}, we have
\[
\cost(\mcQ) \le \cost(\mcP) + \dist(I(\mcP), I(\mcQ)) \le \cost(\mcP) + \dist(I(\mcP), I(\mcO)) \le 2 \cdot \cost(\mcP) + \cost(\mcO).
\]
If the performance guarantees are expected, i.e., $\E[\cost(\mcP)] \le \alpha \cdot \OPT_\infty$,
then taking expectations and using $\OPT_\infty \le \OPT_k$ proves the theorem.

For the running time, one sees that obtaining $\mcQ = \Split_k(\mcP)$ from $\mcP$ can be done in $O(|V|^2)$ time.
\end{proof}

\begin{corollary}
\label{coro01}
For every fixed $\epsilon > 0$, the approximation algorithm of Garc\'ia-Soriano and Schohn~\cite{GS26} for the uncapacitated cluster editing problem
together with the downsizing procedure gives a $(3.7730 + \epsilon)$-approximation algorithm for the capacitated problem.
\end{corollary}
\begin{proof}
The algorithm of Garc\'ia-Soriano and Schohn~\cite[Corollary 1.4]{GS26} can be set to have an approximation ratio of $1.3865 + \epsilon/2$,
and runs in $O(2^{\poly(1/\epsilon)} |V|^{O(1)})$ time.
The resulting algorithm, by \Cref{thm02}, has an approximation ratio of $2 \cdot (1.3865 + \epsilon/2) + 1 = 3.7730 + \epsilon$.
\end{proof}

\section{An improved approximation algorithm}\label{sec:hybrid}
Let $\mcF$ be the collection of all non-empty subsets of $V$.
Replacing $\mcF_k$ in the Cluster-LP in Eq.~(\ref{eq03}) by $\mcF$ becomes the Cluster-LP for the uncapacitated cluster editing problem,
which is employed in the works by Cao et al.~\cite{CCL24} and Garc\'ia-Soriano and Schohn~\cite{GS26} that contain the state-of-the-art approximation algorithms.
Most notably, it is shown that for any rational $0 < \delta \le \frac 12$,
one can deterministically compute a feasible solution $z$ to the Cluster-LP in time $|V|^{\poly(1/\delta)}$ times a polynomial in the precision encoding length,
such that $z$ has at most $|V|$ positive values all of polynomial encoding length and $L(z) \le (1 + \delta) L^*$.

One sees that by setting $z_S = 0$ for each $S \in \mcF \setminus \mcF_k$ in the Cluster-LP for the uncapacitated cluster editing problem,
it reduces to the Cluster-LP in Eq.~\eqref{eq03} for the capacitated problem.
Therefore, we can use the conclusion for the uncapacitated problem and summarize it in the below theorem.

\begin{theorem}[Approximate fractional solution]
\label{thm04}
Given a complete edge-signed graph $G = (V, E^+ \cup E^-)$, an input capacity $k$, and a rational $0 < \delta \le \frac 12$,
one can deterministically compute a feasible solution $z$ to the Cluster-LP in Eq.~(\ref{eq03}) in time $|V|^{\poly(1/\delta)}$ times a polynomial in the precision encoding length,
such that $z$ has at most $|V|$ positive values all of polynomial encoding length and $L(z) \le (1 + \delta) L^*_k$.
\end{theorem}

In the rest of the section, we present a hybrid rounding algorithm for $z$, in Algorithm~\ref{alg01}, and prove the rounding guarantee independently of the LP solver for obtaining $z$.
Inside the algorithm $z$ is used for constructing two partitions.

\begin{algorithm}[htbp]
\caption{Hybrid rounding a fractional solution.\label{alg01}}
\begin{algorithmic}[1]
\Require
	$G = (V, E^+ \cup E^-)$, $k$, a feasible solution $z$ to Cluster-LP in Eq.~\eqref{eq03};
\Ensure
	a partition for $V$.
\State {$U \leftarrow V, \ \mcR \leftarrow \emptyset$;}\Comment{Sampling branch;}
\While {($U \ne \emptyset$)}
	\State {$\mcJ_U = \{S : z_S > 0, S \cap U \ne \emptyset\}$ and $Z_U = \sum_{S \in \mcJ_U} z_S$;}
	\State {choose $S \in \mcJ_U$ with probability $z_S/Z_U$;}
	\State {add $S \cap U$ to $\mcR$ and update $U \gets U \setminus S$;}
\EndWhile
\State {$U \leftarrow V, \ \mcP \leftarrow \emptyset$;}\Comment{Ordinary pivoting branch;}
\While {($U \ne \emptyset$)}
	\State {choose $p \in U$ with probability $1/|U|$;}
	\State {add $C \gets N^+[p] \cap U$ to $\mcP$ and update $U \gets U \setminus C$;}
\EndWhile
\State {\textbf{return} \ $\mcH \gets \arg\min_{\mcQ \in \{\mcR, \Split_k(\mcP)\}} \cost(\mcQ)$.}
\end{algorithmic}
\end{algorithm}

A high-level description of the algorithm is depicted in Algorithm~\ref{alg01}, which constructs two candidate partitions and returns the better one as the final solution.
Lines 1--6 constitute the first {\em sampling} branch that generates a candidate $k$-feasible partition $\mcR$, which is initialized empty at the beginning.
In each iteration, let $U$ denote the un-clustered vertices, and $\mcJ_U$ denote the collection of subsets $S$ such that $z_S > 0$ and $S$ contains at least one un-clustered vertex;
then each $S$ of $\mcJ_U$ is selected with probability proportional to its $z_S$ value and then $S \cap U$ is added to $\mcR$.
The branch repeats until every vertex is in a cluster of $\mcR$.
Lines 7--11 constitute the second branch that is the {\em ordinary pivot} algorithm by Ailon et al.~\cite{ACN08}.
It generates an uncapacitated partition $\mcP$, which is initialized empty at the beginning.
In each iteration, let $U$ denote the un-clustered vertices;
each vertex $p \in U$ is selected uniformly at random as the {\em pivot} to form a cluster $N^+[p] \cap U$, i.e., containing $p$ and all its positive un-clustered neighbors.
The branch repeats until every vertex is in a cluster of $\mcP$, of which $\Split_k(\mcP)$ is the generated candidate $k$-feasible partition.

One sees that either branch has at most $|V|$ iterations, and the $k$-feasibility of the partition is guaranteed.
The overall running time of Algorithm~\ref{alg01} is in $O(|V|^3 \log |V|)$, using independent random bits in the iterations.

\subsection{The performance analysis}
Recall in Eq.~\eqref{eq04} that we define $y_{uv} = \sum_{S \in \mcC(u, v)} z_S$ for each edge $uv$,
where $\mcC(u, v)$ is the collection of subsets of $\mcF_k$ each containing both end-vertices $u, v$.
With respect to the fractional solution $z$, we call $y_{uv}$ the {\em total weight} of the clusters containing both end-vertices,
and call $x_{uv} = 1 - y_{uv}$ the total weight of the clusters containing exactly one of the two end-vertices.
When we fix the edge $uv$ for discussion below, we denote $x := x_{uv}$.
In a $k$-feasible partition, we say $u, v$ are {\em together} if they are in the same cluster of the partition, or otherwise {\em separated}.
The next lemma on the partition $\mcR$ produced in the first branch of Algorithm~\ref{alg01} is analogous to the pairwise sampling principle of Cao et al.~\cite{CCL24}.

\begin{lemma}
\label{lemma02}
In the partition $\mcR$ produced in the first branch of Algorithm~\ref{alg01}, for every pair of vertices $u, v$,
\[
\Pr[u,v \mbox{ together in } \mcR] = \frac {1-x}{1+x},\qquad
\Pr[u,v \mbox{ separated in } \mcR] = \frac {2x}{1+x}.
\]
\end{lemma}
\begin{proof}
Note that $u, v$ are together or separated in $\mcR$ is independent of the cluster construction history before any of $u$ or $v$ is clustered,
at that time point every subset of $\mcF_k$ containing either of them still intersects the current un-clustered vertex set $U$, and thus $x$ has not changed.
That is, for any history the probabilities that the next selected cluster contains both $u$ and $v$, or only $u$, or only $v$ are
$\frac {1-x}{Z_U}$, $\frac x{Z_U}$, $\frac x{Z_U}$, respectively.
It follows that the probabilities condition on one of them being clustered are $\frac {1-x}{1+x}$, $\frac x{1+x}$, and $\frac x{1+x}$, respectively, independent of the history.
If the selected subset $S$ contains both $u$ and $v$, then they are in the same cluster $S \cap U$.
If $S$ contains exactly one of $u$ and $v$, then that vertex is in $S \cap U$ while the other vertex appears in a different cluster, i.e., $u$ and $v$ are separated.
This proves the lemma.
\end{proof}

Recall that for each positive edge $uv$, it contributes $x$ towards the objective in the fractional solution $z$.
By \Cref{lemma02}, we define the {\em slack} of the edge $uv$, denoted as $d_{uv}$,
as the difference between twice of its contribution towards the objective in the fractional solution $z$, $2x_{uv}$,
and its expected contribution towards the objective in the partition $\mcR$, $\frac {2x}{1+x}$.
That is, $d_{uv} = 2 x_{uv} - \frac {2x_{uv}}{1 + x_{uv}} = \frac {2x_{uv}^2}{1+x_{uv}}$.
Analogously, we can define the slack for a negative edge.
We have
\begin{equation}
\label{eq06}
d_{uv} = \begin{cases}
	\dfrac {2x_{uv}^2}{1+x_{uv}},						&uv\in E^+;\\
	\dfrac {(1-x_{uv})(1+2x_{uv})}{1+x_{uv}},	&uv\in E^-;
	\end{cases}
\end{equation}
and $D = \sum_{uv \in E^+ \cup E^-} d_{uv}$ is called the {\em total slack},
which measures exactly the total improvement of the partition $\mcR$ over the bound $2 L(z)$ and is summarized in the next lemma.
One sees that every edge slack $d_{uv}$ is nonnegative.

\begin{lemma}
\label{lemma03}
For the partition $\mcR$ produced in the first branch of Algorithm~\ref{alg01}, $\E[\cost(\mcR)] = 2 L(z) - D$.
\end{lemma}
\begin{proof}
For each edge $uv$, its slack is defined as the difference between twice of its contribution towards the objective in the fractional solution $z$
and its expected contribution towards the objective in the partition $\mcR$.
Summing over all edges gives $D = 2 L(z) - \E[\cost(\mcR)]$, and thus proves the lemma.
\end{proof}

We next use the total slack $D$ to bound the disagreement cost of $\mcP$ produced in the second branch of Algorithm~\ref{alg01}.
Noting that for the selected pivot $p$ in an iteration and a positive neighbor $u \in N^+[p]$,
bringing in another positive neighbor $v \in N^+[p]$ into the cluster $N^+[p] \cap U$ has the risk of introducing an error negative edge $uv$;
while not bringing in a negative neighbor $v \notin N^+[p]$ into the cluster $N^+[p] \cap U$ also has the risk of introducing an error positive edge $uv$.
Both scenarios involve a {\em bad triangle}, which has exactly two positive edges and one negative edge, and plays an important role in the performance analysis.

\begin{lemma}
\label{lemma04}
For every bad triangle $T$, $\sum_{e \in I(T)} d_e \ge 2/3$, where $I(T)$ is the edge set of $T$.
\end{lemma}
\begin{proof}
Assume the three vertices in $T$ are $u, v, w$, and $uv$ is a negative edge.
Let $a, b$ denote the two values $x_{uw}$ and $x_{vw}$ for the positive edges and $c = x_{uv}$.
By the triangle inequality for $x$ in Eq.~\eqref{eq05}, $a + b \ge c$.

The slack function $f(t) = 2t^2/(1+t)$ for positive edges is nondecreasing and convex in $[0, 1]$, since
\[
f'(t) = 2 - \frac 2{(1+t)^2} \ge 0, \mbox{ and } f''(t) = \frac 4{(1+t)^3} > 0.
\]
Therefore, $f(a) + f(b) \ge 2 \cdot f((a+b)/2) \ge 2 \cdot f(c/2)$.
It follows that
\[
\sum_{e \in I(T)} d_e \ge 2 \cdot f(c/2) + \frac {(1-c)(1+2c)}{1+c} = \frac {2c^2}{2+c} + \frac {(1-c)(1+2c)}{1+c}
	= \frac 23 + \frac {(1-c)(5c+2)}{3(c+1)(c+2)} \ge \frac23,
\]
where the last inequality holds due to $0 \le c \le 1$.
\end{proof}

\begin{lemma}
\label{lemma05}
For the uncapacitated partition $\mcP$ produced in the second branch of Algorithm~\ref{alg01}, $\E[\cost(\mcP)] \le \tfrac 92 D$.
\end{lemma}
\begin{proof}
Consider a cluster $C$ added to $\mcP$ at the end of the iteration, in which $U$ is the remaining un-clustered vertex set at the beginning.
Recall that each vertex of $U$ is selected as the pivot $p$ with an equal probability.
We write $C_p = N^+[p] \cap U$ the cluster formed using $p$.
Denote
\begin{align*}
a_p &= e^-(C_p) + e^+(C_p, U \setminus C_p),\\
b_p &= \sum_{\{u, v\} \subseteq U, \{u, v\} \cap C_p \ne \emptyset} d_{uv}.
\end{align*}
One sees that $a_p$ counts the number of edge errors decided in this iteration if $p$ is selected,
and $b_p$ is the total edge slack removed from the remaining instance and thus attributed to $C_p$.

Consider a negative edge error $uv$ introduced by $C_p$:
We have $p \ne u, v$, and thus $pu$ and $pv$ are positive edges and then these three vertices form a bad triangle.

Consider a positive edge error $uv$ introduced by $C_p$:
We assume without loss of generality that $pu$ is positive and thus $pv$ is negative, and then these three vertices form a bad triangle too.

We conclude that for each edge error $uv$ counted towards $a_p$, it determines the unique bad triangle on $\{p, u, v\}$.
Conversely, for every bad triangle containing positive edges $pu, pv$ and negative edge $uv$,
it will be discovered when $p$ ($u, v$, respectively) is the pivot and the edge $uv$ ($pv, pu$, respectively) is the edge error.
In other words, every bad triangle contained in the subgraph induced by $U$ contributes exactly $3$ edge errors when its three vertices are selected as the pivot.

Let $T_U$ be the number of bad triangles contained in the subgraph induced by $U$.
We have 
\begin{equation}
\label{eq07}
\sum_{p \in U} a_p = 3 T_U.
\end{equation}

Also for every bad triangle on $\{p, u, v\}$, the slack $d_{uv}$ of the edge $uv$ is counted towards $b_p$ since at least one of $u, v$ is in $C_p$.
Therefore, the total slack of the edges of the triangle is included in $\sum_{p \in U} b_p$, i.e.,
\begin{equation}
\label{eq08}
\sum_{p \in U} b_p \ge \sum_{T \subseteq U, T \mbox{ is bad}} \sum_{e \in I(T)} d_e \ge \tfrac 23 T_U,
\end{equation}
where the last inequality is by \Cref{lemma04}.

Let $\E[\cost(\mcP|_U)]$ denote the expected cost of the partition $\mcP$ restricted on $U$,
and $D_U$ denote the total slack of the edges in the subgraph induced by $U$.
We prove by induction that $\E[\cost(\mcP|_U)] \le \tfrac 92 D_U$, for which the equality holds trivially when $|U| \le 1$.
When $|U| \ge 2$, we have
\begin{align*}
\E[\cost(\mcP|_U)] &= \frac 1{|U|} \sum_{p \in U} \left( a_p + \E[\cost(\mcP|_{U \setminus C_p})] \right)\\
	&\le \frac 1{|U|} \sum_{p \in U} \left( a_p + \frac 92 D_{U \setminus C_p} \right)\tag{by inductive hypothesis}\\
	&= \frac 1{|U|} \sum_{p \in U} \left( a_p + \frac 92 \left( D_U - b_p \right) \right)\\
	&= \frac 92 D_U + \frac 1{|U|} \sum_{p \in U} \left( a_p - \frac 92 b_p \right)\\
	&\le \frac 92 D_U.\tag{by Eqs.~(\ref{eq07}, \ref{eq08})}
\end{align*}
We then take $U = V$, and thus prove the lemma.
\end{proof}

\begin{theorem}
\label{thm05}
Given a fractional solution $z$ to the Cluster-LP in Eq.~\eqref{eq03},
the hybrid rounding Algorithm~\ref{alg01} runs in expected polynomial time in $|V|$, the number of positive variables in $z$, and the encoding length of $z$,
and returns a $k$-feasible partition $\mcH$ satisfying
\[
\E[\cost(\mcH)] \le \frac 95 L(z) + \frac 1{10} \OPT_k.
\]
\end{theorem}
\begin{proof}
For the running time, suppose $z$ has $m$ positive variables and each of them has at most $b$ bits.
Each iteration of the first and the second branches takes an expected polynomial time in $|V|, m, b$ for probability generation followed by cluster formation.
There are at most $|V|$ iterations in each branch, and thus the overall running time is in expected polynomial time.

For the quality of the returned partition $\mcH$,
by Theorem~\ref{thm02} and Lemma~\ref{lemma05} we have
\[
\E[\cost(\mcQ)] \le 2 \cdot \frac 92 D + \OPT_k = 9D + \OPT_k.
\]
Hence, using Lemma~\ref{lemma03}, we have
\[
\E[\cost(\mcH)] \le \frac 9{10} \E[\cost(\mcR)] + \frac 1{10} \E[\cost(\mcQ)] \le \frac 9{10}(2L(z) - D) + \frac 1{10}(9D + \OPT_k)
	= \frac 95L(z) + \frac 1{10}\OPT_k.
\]
This proves the theorem.
\end{proof}

We are ready to present the complete algorithm in \Cref{alg02} for the capacitated cluster editing problem,
of which an instance consists of a complete edge-signed graph $G = (V, E^+ \cup E^-)$, a capacity $k$, and an approximation tolerance $\epsilon$.

\begin{algorithm}[htbp]
\caption{The complete approximation algorithm.\label{alg02}}
\begin{algorithmic}[1]
\Require
	$G = (V, E^+ \cup E^-)$, $k \ge 1$, a positive rational $\epsilon$;
\Ensure
	a $k$-feasible partition for $V$.
\If {($k = 1$)}
	\State {\textbf{return}} the partition of all singletons;
\ElsIf {(all positive components are cliques of size at most $k$)}
	\State {\textbf{return}} the partition of all the components;
\EndIf
\State {$\delta \leftarrow \min\{\frac 59 \epsilon, \frac 12\}$;}
\State {use \Cref{thm04} to compute a solution $z$;}
\State {run \Cref{alg01} on $z$ to compute a solution $\mcH$;}
\State {\textbf{return} \ $\mcH$.}
\end{algorithmic}
\end{algorithm}

\begin{theorem}
\label{thm01}
For every fixed rational $\epsilon > 0$, the Capacitated Cluster Editing problem admits a randomized expected $(1.9 + \epsilon)$-approximation algorithm
that runs in an expected polynomial time in the input size $n$, bounded by $n^{\poly(1/\epsilon)}$ in which the exponent does not depend on the capacity $k$,
times a polynomial in the precision encoding length.
\end{theorem}
\begin{proof}
The first conditional statement (Lines 1--5) of Algorithm~\ref{alg02} is exact, that gives an optimal solution.
Otherwise, Theorems~\ref{thm04} and~\ref{thm05}, together with $L_k^* \le \OPT_k$, give
\[
\E[\cost(\mcH)] \le \left( \frac 95 \left(1 + \delta \right) + \frac 1{10} \right) \OPT_k \le \left( \frac {19}{10} + \epsilon \right) \OPT_k.
\]
The overall running time of the algorithm follows from the running time stated in Theorems~\ref{thm04} and~\ref{thm05}, for computing $z$ and $\mcH$, respectively.
\end{proof}

\section{An improved pivoting algorithm}\label{mp:section}
In the last section, we adopt the ordinary pivot algorithm of Ailon et al.~\cite{ACN08} to first generate an uncapacitated partition $\mcP$,
independently from the fractional solution $z$ stated in \Cref{thm04},
followed by the splitting procedure to downsize large parts so that a candidate $k$-feasible partition is achieved.
Both the algorithmic steps and their performance analysis appear straight, with the most important guarantee $\E[\cost(\mcP)] \le \frac 92 D$,
where $D = 2 L(z) - \E[\cost(\mcR)]$ is the total edge slack and $\mcR$ is the $k$-feasible partition produced by the first sampling branch of Cao et al.~\cite{CCL24}.

In this section, we propose an improved pivoting algorithm to generate directly a $k$-feasible partition using the fractional solution $z$ stated in \Cref{thm04},
that replaces the ordinary pivot algorithm in \Cref{alg01}.
We continue to use the notations defined in the previous sections,
such as the quantities $y_{uv} = \sum_{S \in \mcC(u, v)} z_S$, $x_{uv} = 1 - y_{uv}$, $d_{uv}$ associated with an edge $uv$, etc.
Nevertheless, the upcoming algorithmic description and the analysis use mostly $y_{uv}$'s.
We therefore write the edge slack $d_{uv}$ in Eq.~\eqref{eq06} as an equivalent function of $y_{uv}$ below.

Let
\[
f_+(y) = \frac {2(1-y)^2}{2-y}, \qquad
f_-(y) = \frac {y(3-2y)}{2-y};
\]
the slack of edge $uv$ in Eq.~\eqref{eq06} can be re-written as
\begin{equation}
\label{eq09}
d_{uv} = \begin{cases}
	f_+(y_{uv}), &uv \in E^+;\\
	f_-(y_{uv}),	&uv \in E^-.
	\end{cases}
\end{equation}

Given a partition $\mcP$, when it is not $k$-feasible, we have proven in \Cref{thm02} that
downsizing every part of $\mcP$ of size greater than $k$ by splitting it into a number of parts of size exactly $k$ and a possible residual part of size less than $k$,
leads to a $k$-feasible partition $\Split_k(\mcP)$ satisfying $\cost(\Split_k(\mcP)) \le 2 \cdot \cost(\mcP) + \OPT_k$.
The proof of this inequality uses mostly the triangle inequality of the edge set edit distance.
Below we separate the disagreement cost of $\mcP$ into two parts and prove a more precise upper bound.

We denote $p(\mcP) = |E^+ \setminus I(\mcP)|$ the number of positive edge errors and $n(\mcP) = |I(\mcP) \setminus E^+| = |E^- \cap I(\mcP)|$ the number of negative edge errors.
Clearly, $\cost(\mcP) = p(\mcP) + n(\mcP)$.

\begin{lemma}
\label{lemma06}
For every partition $\mcP$, the downsized $k$-feasible partition $\Split_k(\mcP)$ satisfies
\[
\cost(\Split_k(\mcP)) \le p(\mcP) + 2 n(\mcP) + \OPT_k.
\]
\end{lemma}
\begin{proof}
Let $F = |I(\mcP) \setminus I(\Split_k(\mcP))|$ denote the number of edges deleted during the downsizing process,
and let $t^-$ be the number of negative edges among these $F$ edges.

By \Cref{lemma01}, for an optimal $k$-feasible partition $\mcO$, we have $F \le |I(\mcP) \setminus I(\mcO)| \le \OPT_k + n(\mcP)$,
since every positive edge of $I(\mcP) \setminus I(\mcO)$ is an error of $\mcO$ and every negative edge is an error counted towards $n(\mcP)$.
Since the downsizing process changes the cost by $(F - t^-) - t^-$, and $t^- \ge 0$, we have
\[
\cost(\Split_k(\mcP)) = \cost(\mcP) + F - 2 t^- \le (p(\mcP) + n(\mcP)) + (\OPT_k + n(\mcP)) = p(\mcP) + 2 n(\mcP) + \OPT_k.
\]
This proves the lemma.
\end{proof}

\Cref{lemma06} together with the edge slacks in Eq.~\eqref{eq09} suggest that
the second pivoting branch can have better trade-off between positive and negative neighbors when forming a cluster,
and that the positive and negative edge errors can be analyzed with unequal weights.

\subsection{The modified pivoting branch}\label{mp:algorithm-section}
Our modified pivot rounding algorithm follows the conditional seed-based architecture of Garc\'ia-Soriano and Schohn~\cite{GS26},
where a seed is the selected subset containing the chosen pivot and,
while positive neighbors inside the seed are surely admitted into the cluster, a different admission rule is set for positive neighbors outside the seed.
To this purpose, we set up
\begin{equation}
\label{eq10}
\beta = \frac {153}{50}, \qquad
	r(y) = \max\left\{ 0, 1 - \beta \frac {1-y}{2-y}\right\}, \qquad
	m(y) = \min\left\{ \frac 32 y, 2y - y^2 \right\}.
\end{equation}
One sees that both $r(y)$ and $m(y)$ are increasing in $y$.
The modified pivot rounding algorithm takes in a fractional solution $z$ to Cluster-LP in Eq.~\eqref{eq03} and returns an uncapacitated partition $\mcP$.
The final produced candidate $k$-feasible partition is $\Split_k(\mcP)$.

In the algorithm, assume $U$ is the current set of un-clustered vertices.
It chooses a pivot $p$ from $U$ uniformly at random, and then select the seed $S$ from all those subsets containing $p$, i.e., $\mcC(p)$, with probability $z_S$
(recall the vertex coverage equality in Eq.~\eqref{eq03} that $\sum_{S \in \mcC(p)} z_S = 1$).
The to-be constructed cluster $C$ initially contains $p$ and all its positive un-clustered neighbors in $S$, i.e., $N^+[p] \cap (U \cap S)$;
then for each positive neighbor $v \in U \setminus S$, it is added to $C$ only if $y_{pv} \ge \frac {53}{103}$ and with probability $r(y_{pv})$.
Every un-clustered negative neighbor $v$ is treated the same, that it is added to $C$ with probability $m(y_{pv})$.
This forms a new cluster $C$ in $\mcP$ and the algorithm repeats until all vertices are clustered.
A pseudocode of this modified pivot rounding is depicted in \Cref{alg03}.

\begin{algorithm}[htbp]
\caption{Modified pivot rounding a fractional solution.\label{alg03}}
\begin{algorithmic}[1]
\Require
	$G = (V, E^+ \cup E^-)$, $k$, a feasible solution $z$ to Cluster-LP in Eq.~\eqref{eq03};
\Ensure
	an uncapacitated partition $\mcP$ for $V$.
\State {$U \leftarrow V, \ \mcP \leftarrow \emptyset$;}
\While {($U \ne \emptyset$)}
	\State {choose $p \in U$ with probability $1/|U|$;}\Comment{$p$ the pivot;}
	\State choose $S \in \mcC(p)$ with probability $z_S$;
	\State $C \gets \{p\}$;
	\For {(each $v \in U \setminus \{p\}$)}
		\If {($pv \in E^+$ and $v \in S$)}
			\State $C \gets C \cup \{v\}$;\Comment{$v$ positive neighbor inside $S$;}
		\ElsIf {($pv \in E^+$)}
			\State $C \gets C \cup \{v\}$ with probability $r(y_{pv})$;\Comment{$v$ positive neighbor outside $S$;}
		\Else
			\State $C \gets C \cup \{v\}$ with probability $m(y_{pv})$;\Comment{$v$ negative neighbor, inside or outside $S$;}
		\EndIf
	\EndFor
	\State {add $C$ to $\mcP$ and update $U \gets U \setminus C$;}
\EndWhile
\State {\textbf{return} \ $\mcP$.}
\end{algorithmic}
\end{algorithm}

\subsection{A weighted performance analysis}\label{mp:analysis}
Instead of the disagreement cost of $\mcP$, $\cost(\mcP) = p(\mcP) + n(\mcP)$ as in \Cref{lemma05}, in this section we estimate the quantity $p(\mcP) + 2 n(\mcP)$,
which can be deemed as {\em weighted} cost of $\mcP$ by assigning twice weight to a negative edge of $\mcP$.
We prove the following lemma.

\begin{lemma}
\label{lemma07}
The partition $\mcP$ produced by \Cref{alg03} satisfies $\E[p(\mcP)+2n(\mcP)]\le\beta D$.
\end{lemma}
\begin{proof}
Consider a cluster $C$ added to $\mcP$ at the end of the current iteration, which is a subset of $U$ the remaining un-clustered vertices in the iteration.
Note that each vertex of $U$ is selected as the pivot $p$ with an equal probability;
we write $C_p$ for the random cluster for $p$ and denote
\[
a_p = 2 e^-(C_p) + e^+(C_p, U \setminus C_p),\qquad
b_p = \sum_{{\{u, v\} \subseteq U, \ \{u, v\} \cap C_p \ne \emptyset}} d_{uv}.
\]
Similar to the proof of \Cref{lemma05}, $e^-(C_p)$ counts negative edge errors, $e^+(C_p, U \setminus C_p)$ counts positive edge errors,
but this time $a_p$ is the {\em weighted} total error decided in this iteration.
Note that $b_p$ is kept the same, which is the total edge slack removed from the remaining instance and thus attributed to the cluster $C_p$.

We prove the following inequality
\begin{equation}
\label{eq11}
\sum_{p \in U} \E[a_p] \le \beta \sum_{p \in U} \E[b_p],
\end{equation}
where the expectations are conditional on the fixed history leading to the beginning of this current iteration.

From how a positive neighbor is admitted into the cluster,
we see that a positive edge $pv$ is included in $I(\mcP)$ with probability $\mu_{pv} = y_{pv} + (1 - y_{pv}) r(y_{pv}) \triangleq M(y_{pv})$,
and thus excluded with probability
\begin{equation}
\label{eq12}
1 - \mu_{pv} = 1 - y_{pv} - (1 - y_{pv}) \cdot \max\left\{ 0, 1 - \beta \frac {1-y_{pv}}{2-y_{pv}}\right\} = \min\left\{1-y_{pv}, \frac \beta2 f_+(y_{pv})\right\}.
\end{equation}
In particular, when $y_{pv} \ge \frac {53}{103}$, the edge exclusion probability is exactly $\frac \beta2 f_+(y_{pv})$.

On the other hand, a negative edge $pv$ is included in $I(\mcP)$ with probability $\mu_{pv} = m(y_{pv})$.

Similar to the proof of \Cref{lemma05}, we use triangles to prove the joint probabilities of the edges of $I(\mcP)$.
We denote $\gamma_{pv} = 1 - r(y_{pv})$ if the edge $pv$ is positive, or otherwise $\gamma_{pv} = 0$.
Consider a triangle $T$ on three distinct vertices $p, u, v$, and let $\mcC(p, u, v) = \{S \in \mcF_k: \{p, u, v\} \subseteq S\}$, followed by $t_{puv} = \sum_{S \in \mcC(p, u, v)} z_S$.
With the selected pivot $p$, the seed $S \in \mcC(p, u, v)$
($\mcC(p, u) \setminus \mcC(p, u, v), \mcC(p, v) \setminus \mcC(p, u, v), \mcC(p) \setminus (\mcC(p, u) \cup \mcC(p, v))$, respectively)
has probability $t$ ($y_{pu} - t$, $y_{pv} - t$, $1 - y_{pu} - y_{pv} + t$, respectively), where $t \triangleq t_{puv}$.
Therefore, the {\em joint probability} that both $u, v$ are included in $C_p$~\cite{GS26} is
\begin{align}
J_{p;uv} &= \begin{cases}
	t + (y_{pu} - t) r(y_{pv}) + (y_{pv} - t) r(y_{pu}) + (1 - y_{pu} - y_{pv} + t) r(y_{pu}) r(y_{pv}),	& u, v \mbox{ both positive};\\
	\mu_{pu} \mu_{pv},	&\mbox{ otherwise}.
	\end{cases}\nonumber\\
	&= \mu_{pu} \mu_{pv} + \gamma_{pu} \gamma_{pv} (t_{puv} - y_{pu} y_{pv}).\label{eq13}
\end{align}

It follows that the weighted error made by the edge $uv$ is
\begin{equation}
\label{eq14}
A_{p;uv} = \begin{cases}
	\mu_{pu} + \mu_{pv} - 2 J_{p;uv}, 	&uv \in E^+,\\
	2 J_{p;uv},										&uv \in E^-.
	\end{cases}
\end{equation}
For $uv$, as long as one endpoint is included in the cluster $C_p$, its slack is attributed to $C_p$,
and one sees that such probability is $\mu_{pu} + \mu_{pv} - J_{p;uv}$;
the slack contributed by $uv$ is thus
\begin{equation}
\label{eq15}
B_{p;uv} = d_{uv} (\mu_{pu} + \mu_{pv} - J_{p;uv}).
\end{equation}
Since every vertex of the triangle $T$ can be chosen as the pivot in the iteration, we denote
\[
A_T = A_{p;uv} + A_{u;vp} + A_{v;pu}, \qquad B_T = B_{p;uv} + B_{u;vp} + B_{v;pu},
\]
and we will validate that $A_T \le \beta B_T$ for every triangle $T$.

\paragraph{Triangle validation.}
We use Eqs.~\eqref{eq13}--\eqref{eq15} to validate $A_T \le \beta B_T$ for all four kinds of triangles $T$, which have $3$, $2$, $1$ and $0$ positive edges
(denoted as $+^3$-, $+^2$-, $+^1$- and $+^0$-triangle), respectively.
Let $a, b, c$ denote the three unordered values $y_{pu}, y_{pv}, y_{uv}$.
Recall that $t = t_{puv} = \sum_{S \in \mcC(p, u, v)} z_S$.
From the vertex coverage equalities of Eq.~\eqref{eq03}, we have 
\begin{equation}
\label{eq16}
\max\{ 0, a+b-1, a+c-1, b+c-1 \} \le t \le \min\{ a, b, c \}.
\end{equation}

For a $+^3$-triangle $T$, order $a \le b \le c$;
Eq.~\eqref{eq16} becomes $b + c - 1 \le t \le a$, and $\beta B_T - A_T$ is a linear function in $t$.
It suffices to check that $\beta B_T - A_T \ge 0$ at $t = a$ and $t = \max\{ 0, b+c-1 \}$.

For a $+^2$-triangle $T$, let $a \le b$ be the values for two positive edges and $c$ the value for the negative edge;
Eq.~\eqref{eq16} becomes $a + b - 1 \le t \le c$ and $b + c - 1 \le t \le a$.
The derivative of $\beta B_T - A_T$ with respect to $t$ is $- (\beta f_-(c) + 2)(1 - r(a))(1 - r(b)) \le 0$;
hence it suffices to check that $\beta B_T - A_T \ge 0$ at $t = \min\{a, c\}$.

For a $+^1$-triangle, let $a$ be the value for the positive edges and $b \le c$ be the values for the two negative edges;
Eq.~\eqref{eq16} becomes $a + c - 1 \le b$ and $b + c - 1 \le a$, since one sees that $t$ does not appear in $\beta B_T - A_T$.

For each of the above three kinds of triangles,
the function $\Psi(a, b, c) \triangleq (2 - a)(2 - b)(2 - c)(\beta B_T - A_T)$ is a polynomial and $(2 - a)(2 - b)(2 - c) > 0$ on $[0, 1]^3$.
The non-negativity of the function can be verified by the {\em exact rational certificate}, and thus $A_T \le \beta B_T$ is proved.


Lastly, for a $+^0$-triangle, i.e., all three edges of the triangle are negative, we first have $m(y) \le \frac 98 f_-(y)$ because
\[
\frac 98 f_-(y) - m(y) = \begin{cases}
	\frac {3y(1-2y)}{8(2-y)},				&0 \le y \le \frac 12,\\
	\frac {y(2y-1)(5-4y)}{8(2-y)}, 	&\frac 12 \le y \le 1.
	\end{cases}
\]
Then, $A_T = 2 (m(a) m(b) + m(b) m(c) + m(c) m(a))$ and
\[
B_T \ge \frac 89 (2 (m(a) m(b) + m(b) m(c) + m(c) m(a)) - 3 m(a) m(b) m(c)) \ge \frac 89 (m(a) m(b) + m(b) m(c) + m(c) m(a)).
\]
Consequently, $A_T \le \frac 94 B_T \le \beta B_T$.
This proves $A_T \le \beta B_T$ for the last kind of triangles.

It remains to consider the last case where $|U| = 2$, in which the pivot is an endpoint of the last remaining edge $uv$.
One sees that either way the slack $d_{uv}$ is attributed to the formed cluster.
By Eq.~\eqref{eq12}, if $uv$ is positive then it contributes a cost of $1 - \mu_{uv} \le \frac {\beta}2 f_+(y_{uv})$;
or otherwise it contributes a weighted cost of $2 m(y_{uv}) \le \frac 94 f_-(y_{uv}) \le \beta f_-(y_{uv})$.
We have thus proved that, when $|U| = 2$, for the last edge $uv$ its weighted cost is no greater than $\beta d_{uv}$.

Adding $A_T \le \beta B_T$ over all triangles and possibly the above inequality for the last edge proves Eq.~\eqref{eq11},
which gives the conditional one-iteration inequality for the uniformly selected pivot $p$ by dividing $|U|$.
Summing the conditional inequalities over at most $|V|$ iterations proves the lemma.
\end{proof}

\subsection{The resulting approximation guarantee}
\begin{theorem}
\label{mp:main}
Given a fractional solution $z$ to the Cluster-LP in Eq.~\eqref{eq03},
the cheaper partition $\mcH$ between the $k$-feasible solution $\mcR$ by the first branch of Algorithm~\ref{alg01} and
the $\Split_k(\mcP)$ where $\mcP$ produced by Algorithm~\ref{alg03} satisfies
\[
\E[\cost(\mcH)] \le \frac {306}{203} L(z) + \frac {50}{203} \OPT_k.
\]
The overall running time is in expected polynomial time in $|V|$, the number of positive values in $z$, and the encoding length of $z$.
\end{theorem}
\begin{proof}
Let $\mcQ = \Split_k(\mcP)$.
Lemmas~\ref{lemma06} and~\ref{lemma07} give $\E[\cost(\mcQ)] \le \beta D + \OPT_k$.
Recall that \Cref{lemma03} guarantees $\E[\cost(\mcR)] \le 2 L(z) - D$.
The cheaper cost is no greater than any linear combination of the two costs, and therefore,
\[
\E[\cost(\mcH)] \le \frac {153}{203} \left(2 L(z) - D\right) + \frac {50}{203} \left(\frac {153}{50} D + \OPT_k \right) = \frac {306}{203} L(z) + \frac {50}{203} \OPT_k.
\]
The running time of the entire algorithm is similar to the proof of \Cref{thm05}.
\end{proof}

Given that a fractional solution $z$ can be computed in polynomial time so that $L(z) \le (1 + \delta) L^*_k$, for any positive $\delta$ (\Cref{thm04}),
we choose $\delta = \min\{ 1/2, \epsilon/2\}$ for any positive $\epsilon$.
Using $L^*_k \le \OPT_k$, \Cref{mp:main} gives
\[
\E[\cost(\mcH)] \le \left( \frac {356}{203} + \epsilon \right) \OPT_k, \mbox{ where } \frac {356}{203} = 1.753694581280\dots.
\]

\section{Conclusion}
We studied the capacitated Cluster Editing problem where a graph is given,
and the goal is to convert it into a collection of cliques of upper bounded size via a minimum number of edge insertions and deletions.
The problem has been studied in the literature but not much, with the best approximation algorithm of ratio $5.37$.
Nevertheless, the uncapacitated variant has received at least two waves of extensive studies with several algorithmic frameworks,
and the state-of-the-art approximation ratio of $1.3865$.

In this paper, we presented three improved approximation algorithms, each having a distinct role.
Our first $(3.7730 + \epsilon)$-approximation algorithm builds on the proposed downsizing procedure,
that converts an uncapacitated partition into a feasible one without increasing the cost too much,
and it is achieved by calling the state-of-the-art $1.3865$-approximation algorithm for uncapacitated problem.
The second algorithm employs the near optimal fractional solution to the Cluster LP formulation of the problem,
and combines the rounding through sampling by Cao et al.~\cite{CCL24} and the ordinary pivoting method by Ailon et al.~\cite{ACN08}.
Using the downsizing procedure again, the second algorithm is proven to have an approximation ratio of $1.9 + \epsilon$.
Replacing the ordinary pivoting method in the second algorithm by the conditional affine pivoting rule of Garc\'ia-Soriano and Schohn~\cite{GS26},
again together with the downsizing procedure, the third algorithm is a further improved $(1.7537 + \epsilon)$-approximation algorithm.
We remark that the downsizing procedure, despite simple, has its important role in all three algorithms.
It is also important to note that we extended the algorithm design and analysis techniques developed for the uncapacitated problem,
though not novel but non-trivially.
It remains interesting to see further development for either uncapacitated or capacitated cluster editing,
and improved lower bounds on the approximation ratios.

\bibliographystyle{plain}

\end{document}